\documentclass[aps,prb,showpacs,reprint,amsmath,amssymb,superscriptaddress,a4paper]{revtex4-1}
\usepackage[utf8]{inputenc}
\usepackage{amsmath}
\usepackage{amsfonts}
\usepackage{amssymb}
\usepackage{graphicx}

\usepackage{graphicx}
\usepackage{dcolumn}
\usepackage{bm}
\usepackage{tcolorbox}
\usepackage{color}
\usepackage{tikz} 
\usepackage{lettrine}
\usepackage{ragged2e}
\usepackage{xr} 
\definecolor{mybrown}{RGB}{190, 100, 10}
\colorlet{mybrowntransparent}{mybrown!10}

\begin{document}

\title{$\mu_\mathrm{2T}(n)$: A Method for Extracting the Density Dependent Mobility in Two-Terminal Nanodevices}
\author{Christian E. N. Petersen}
\affiliation{Department of Energy Conversion and Storage, Technical University of Denmark, Kgs Lyngby, Denmark}
\author{Damon J. Carrad}
\affiliation{Department of Energy Conversion and Storage, Technical University of Denmark, Kgs Lyngby, Denmark}
\author{Thierry Désiré}
\affiliation{Department of Energy Conversion and Storage, Technical University of Denmark, Kgs Lyngby, Denmark}
\author{Daria Beznasyuk}
\affiliation{Department of Energy Conversion and Storage, Technical University of Denmark, Kgs Lyngby, Denmark}
\author{Jung-Hyun Kang}
\affiliation{Center for Quantum Devices, Niels Bohr Institute, University of Copenhagen, Copenhagen, Denmark}
\author{Dāgs Olšteins}
\affiliation{Department of Energy Conversion and Storage, Technical University of Denmark, Kgs Lyngby, Denmark}
\author{Gunjan Nagda}
\affiliation{Department of Energy Conversion and Storage, Technical University of Denmark, Kgs Lyngby, Denmark}
\author{Dennis V. Christensen}
\affiliation{Department of Energy Conversion and Storage, Technical University of Denmark, Kgs Lyngby, Denmark}
\affiliation{Institute for Advanced Study, Technical University of Munich, Lichtenbergstrasse 2a, D-85748 Garching, Germany}
\author{Thomas Sand Jespersen}
\affiliation{Department of Energy Conversion and Storage, Technical University of Denmark, Kgs Lyngby, Denmark}
\affiliation{Center for Quantum Devices, Niels Bohr Institute, University of Copenhagen, Copenhagen, Denmark}
\email{tsaje@dtu.dk}

\date{\today}

\begin{abstract}
Measuring carrier mobility as a function of the carrier density in semiconductors using Hall effect is the gold standard for quantifying scattering mechanisms. However, for nanostructures, the Hall effect is not applicable, and the density dependence of mobility is generally inaccessible, rendering Hall effect measurements impractical.  Here, we present $\mu_\mathrm{2T}(n)$, a new procedure allowing us to extract the density dependent mobility in two-terminal measured nano scale field effect transistors at zero magnetic field from conventional conductance vs gate voltage measurements. We validate $\mu_\mathrm{2T}$ against standard Hall measurements and then apply the procedure to 256 individual two-terminal InAs nanowire FETs, extracting information about the scattering mechanisms. To illustrate its broad utility, we reanalyze published data in which mobility had been treated as density independent. Our method represents a new powerful tool for optimization and development of nanomaterials crucial for a wide range of new technologies.

\end{abstract}

\maketitle

\begin{figure*}
\includegraphics[width=1\textwidth]{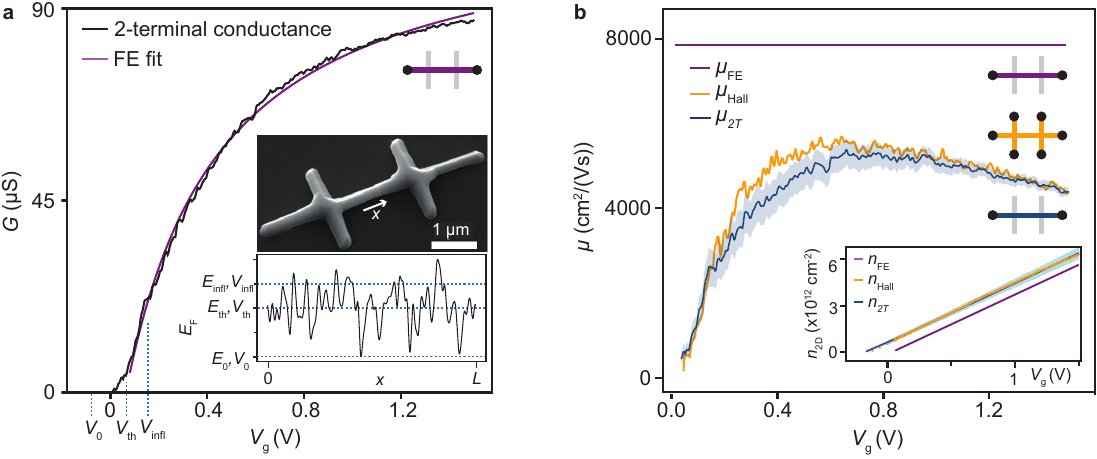}
\caption{\textbf{a}, 2-terminal $G$ vs. $V_\mathrm{g}$ measured of a nanowire device in a Hall bar geometry (see SEM image inset). The purple line is a fit to Eq.~\ref{2point1}. The bottom inset illustrates the random potential landscape due to charged impurities, showing conduction band fluctuations along the nanowire axis ($x$). Dashed lines map critical gate voltages - $V_0$, $V_\mathrm{th}$, and $V_\mathrm{infl}$ - to corresponding energy levels.  
\textbf{b}, Extracted mobilities: mobility $\mu_\mathrm{FE}$ (from the fit in panel a), Hall mobility $\mu_\mathrm{Hall}$ (from Hall measurements), and the two-terminal mobility $\mu_\mathrm{2T}$ (from our analysis method). The inset shows the corresponding carrier densities extracted by each method.
\label{fig:Fig1} }
\end{figure*}

The development and optimization of nanoscale materials and devices is intimately tied to the scientific and technical progress in consumer electronics\cite{delAlamoNat11,caoFutureTransistors2023} and fundamental research\cite{lutchynNRM18,dassarmaNatPhys23,StormerRMP99,ChungNatMat2021}. Bottom-up nanomaterials are set to be central to the scale-up and commercialization of quantum technologies, placing stringent requirements on device reproducibility and stability\cite{neyens_probing_2024,borsoi_shared_2023}, which in turn demands ever-higher materials quality. The most important quality metric is the charge carrier mobility, $\mu$. However, $\mu$ is difficult to determine for low-dimensional nanostructures that are incompatible with the canonical Hall geometries. For example, fabrication of electrodes to access the Hall effect in quasi-one-dimensional nanotubes or nanowires is technically demanding beyond routine\cite{storm:2012,blomers:2012,Hultin2}. Furthermore, Hall electrodes can significantly influence transport in nanostructures, which complicates data interpretation even for quasi-two-dimensional structures\cite{choiAFM2018,gluschkeImpactInvasiveMetal2020a}. 
Terahertz/Raman spectroscopy\cite{BolandNL18}, cathodoluminescence\cite{storm:2012}, and geometric magnetoresistance\cite{OlaussonLind23} can also provide indicative measures of $\mu$; however, without the accuracy to replace transport-based techniques. Due to the simplicity of device fabrication, analysis, and measurement schemes, a two-terminal field effect transistor (FET) is the most common device geometry for semiconductor nanostructures, where the dependence of the conductance, $G$, on a gate-controlled charge carrier density, $n(V_g)$, is analyzed to obtain the field effect mobility $\mu_\mathrm{FE}$\cite{huangNL2002,LindNL2006,gulHighMobilityInSb2015,choiNatMat2018}. However, the use of $\mu_\mathrm{FE}$ as a metric for estimating the material quality is problematic: Most importantly, $\mu_\mathrm{FE}$ is a constant, density independent parameter, while it is well-known that $\mu$ depends strongly on $n$ and in most devices, it is the mobility in a specific density regime which determines performance. Further, the analysis of $\mu(n)$ is the most important tool for understanding the dominating scattering mechanisms limiting $\mu$, such as charged impurities, surface/interface states, and dislocations\cite{dassarmaNatPhys23,makAPL2010,kawaharazukaJAP1999, Ahn_2022, SodergrenIEEE2023}. Such analysis has been key for the highly successful optimization of 2D heterostructures\cite{TsuiPRL82,StormerRMP99,ChungNatMat2021} and is inaccessible through $\mu_\mathrm{FE}$. Finally, even as an estimate of the peak mobility, $\mu_\mathrm{peak}$, $\mu_\mathrm{FE}$ can both over- or underestimate the real value\cite{choiNatMat2018,storm:2012}, thus providing misleading indications of material quality. The inability to routinely access  $\mu(n)$ is thus a serious blind spot in the continuous efforts to optimize nanostructure materials and devices\cite{Ahn_2022,dassarmaNatPhys23}.

Modeling the FET transfer characteristic, $G(V_\mathrm g)$, requires an estimate of the gate capacitance, threshold voltage, and potential contact resistance in series with the device; thus, treating $\mu$ as a $V_\mathrm g$-independent parameter is a natural consequence of concerns with over-fitting. However, $G(V_\mathrm g)$ contains, in principle, information about the full $\mu(n)$ and the main result of the present manuscript is the development of a robust and practical procedure for extracting $\mu(n)$ from conventional two-terminal FET-type measurements, thereby marrying the simplicity of FET device fabrication and measurements with the rich information obtained from Hall characterization. We provide a detailed validation of the results of the new procedure against Hall characterizations enabled through crystal growth of InAs nanowire Hall devices. Subsequently, we demonstrate the use of the model for obtaining the low temperature density dependent mobility from the transfer characteristics of 256 individual two-terminal InAs nanowire FET devices, and we provide a statistical comparison with the results obtained by applying the conventional $\mu_\mathrm{FE}$ analysis. With access to $\mu(n)$, we discuss the dominating scattering mechanisms in InAs nanowires at low temperature and the implications for materials optimization. Finally, to facilitate the adaptation of the procedure, we provide a step-by-step guide, open-access fitting procedures, and an online fitting tool, and we demonstrate 
how transport results from the existing literature for a selection of key material systems can be successfully re-analyzed. This confirms the wide applicability and robustness of the method and illustrates how new information can be obtained from the already existing data across multiple material platforms. Our methods and results enable proper characterization and thus eventual reduction of the disorder in nanoscale materials and devices, which is of crucial importance for the development of solid-state-based quantum technologies.

\section*{Methods for charge carrier mobility estimation}
Figure~1 compares three methods for extracting $\mu$ from devices based on InAs/InGaAs Hall-bar shaped nanostructure. The results obtained by our new procedure described below is denoted $\mu_\mathrm{2T}(n)$ to distinguish from $\mu_\mathrm{FE}$ or $\mu_\mathrm{Hall}(n)$ . The structure, shown in the inset to Fig.~1\textbf{a} before fabrication of contacts and gates, was realized by selective area growth as previously described \cite{KrizekPRM2018,beznasyukPRM22}. Ti/Au Ohmic contacts were fabricated on each of the six arms and a top-gate electrode, isolated from the InAs by 15~nm of HfO$_\mathrm{x}$, which allows tuning of $n$ in the InAs. Details of the fabrication and a scanning electron micrograph of the finished device are included in Supplementary Figure \textbf{S1}. The InAs hosts a two-dimensional electron layer at the surface\cite{blomers:2012,LindNL2006,fordNL09,GuptaNanotech2013,Ahn_2022,KrizekPRM2018,beznasyukPRM22} and Fig.\ 1\textbf{a} shows the two-terminal conductance $G(V_\mathrm{g})$ at $T=20$~mK. The $n$-independent $\mu_\mathrm{FE}$ can be extracted by fitting the conventional Drude expression
\begin{equation}
    G^{-1}(V_\mathrm{g})=R_\mathrm{S}+L^2/(\mu_\mathrm{FE}C(V_\mathrm{g}-V_\mathrm{th}))
    \label{2point1}
\end{equation}
where the constant series resistance $R_\mathrm{S}$ accounts for the metal/semiconductor contact resistance, the resistance of non-gated sections of the device, and the resistance of wiring in the measurement set-up. The gate length, $L=3.8\, \mu \mathrm m$, was measured by electron microscopy, and $C=5.3$~fF is the gate-InAs capacitance obtained by numerical finite element simulations (see Fig. \textbf{S1}$\&$\textbf{S2}). Finally, the fit parameter $V_\mathrm{th}$ is the threshold gate-voltage\cite{gulHighMobilityInSb2015} at which $G$ vanishes. The fit shown in Fig.~1\textbf{a} yields $R_\mathrm{S}= 8.6$~k$\Omega$, $V_\mathrm{th} = 50$~mV and a $n$-independent $\mu_\mathrm{FE} = 7900$~cm$^2$/(Vs),  consistent with previous results on similar materials\cite{blomers:2012,LindNL2006,fordNL09,GuptaNanotech2013,Ahn_2022,KrizekPRM2018,beznasyukPRM22}.


\begin{figure*}
\begin{tcolorbox}[colback=mybrowntransparent,colframe=mybrowntransparent]
\textbf{Box 1. Step-by-step procedure for extracting $n$-dependent 2-terminal mobility, $\mu_\mathrm{2T}(n)$}
\begin{widetext}    

\begin{center}
\includegraphics[width=\textwidth,page=1]{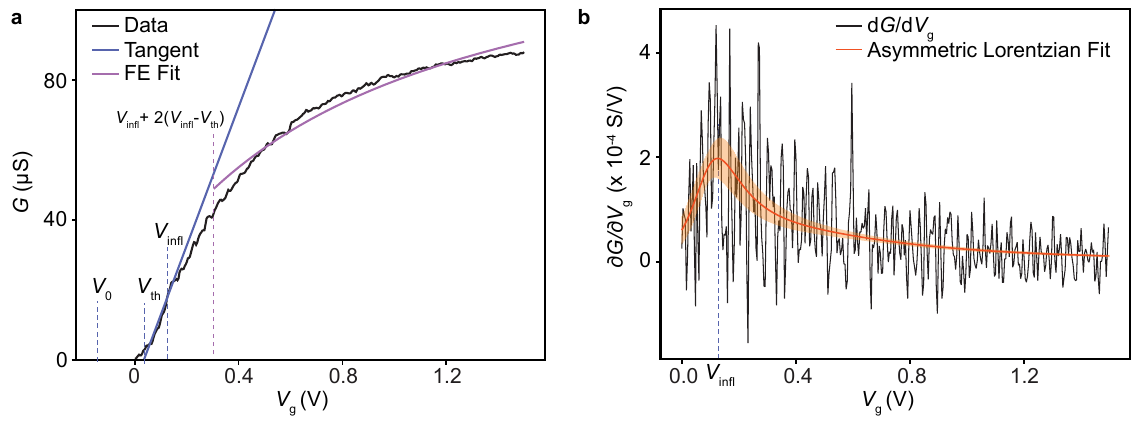} 
\label{Boxfig}
\end{center}

\begin{minipage}[t]{0.48\linewidth} 

\justifying 

Panel \textbf{a} shows a typical measurement of the FET transfer characteristics for a semiconductor nanowire repeated from Fig.\ 1\textbf{a}. To extract the $n$-dependent $\mu$ from this trace, the 5 steps are followed. An online tool is available which implements the procedure for easy extraction of $\mu_\mathrm{2T}(n)$ and $n$ from any two-column $G$ vs $V_\mathrm{g}$ dataset where $C$ and $L$ are known (see Supplementary Section 2). The python source code is also supplied for facile batch processing of large numbers of data set (c.f. Fig.~2)\\

\emph{\textbf{Step 1}: Identify the inflection point, $V_\mathrm{infl}$, and corresponding transconductance $\partial G/\partial V_\mathrm{g}|_\mathrm{max}$}\\

Calculate $\partial G/\partial V_\mathrm{g}$ (panel \textbf{b}) and identify the global maximum, $\partial G/\partial V_\mathrm{g}|_\mathrm{max}$, occurring at $V_\mathrm{infl}$. Smoothing may be required, and as shown in panel \textbf{b} the maximum is identified by fitting $\partial G/\partial V_\mathrm g$ to an asymmetric Lorentzian of the form $f(V_\mathrm G) = 2A \cdot (\pi U_0 (1+4((V_\mathrm g-V_\mathrm{infl})/U_0)^2))^{-1}$ where $U_0 = 2c \cdot (1 + \mathrm{exp}(a(V_\mathrm g-V_\mathrm{infl})))^{-1}$ and $A, c$ and $a$ are fit parameters relating to peak height, width and asymmetry. The uncertainties in panel~\textbf{b} are obtained using error propagation. \\

\emph{\textbf{Step 2}: Identifying threshold voltage, $V_\mathrm{th}$}
\\

Using $V_\mathrm{infl}$ and $\partial G/\partial V_\mathrm{g}|_\mathrm{max}$, extrapolate to the $V_\mathrm{g}$ intercept to define $V_\mathrm{th}$ (panel \textbf{a}). This step is similar to the conventional method for locating $V_\mathrm{th}$\cite{sze_textbook,gulHighMobilityInSb2015}\\

  \

\end{minipage}%
\hfill
\begin{minipage}[t]{0.48\linewidth} 

\justifying 

\emph{\textbf{Step 3}: Identify $V_\mathrm{0}$ and calculate $n(V_\mathrm{g})$.}\\

Estimate the $n=0$ gate voltage, $V_\mathrm{0}$, as $V_\mathrm{0}=V_\mathrm{th}-2(V_\mathrm{infl}-V_\mathrm{th})$ (c.f. Fig.~1\textbf{a}). The density, $n$, is then given by $n(V_\mathrm{g})=C(V_\mathrm{g}-V_\mathrm{0})/(eLW)$ where $C$ is capacitance (see Supplementary Figs. \textbf{S1} $\&$ \textbf{S2}), and $L,W$ the channel length and width. \\
     
\emph{\textbf{Step 4}: Estimate the series resistance, $R_\mathrm{S}$}
\\
       
In the limit $V_\mathrm{g}\gg V_\mathrm{infl}$, the Drude model is approximately valid and can be used to estimate $R_\mathrm{S}$ (panel \textbf{b}) while assuming in this step a constant $\mu$ from
\begin{equation*}
G_\mathrm{FE}^{-1}(V_\mathrm{g}) = R_\mathrm{S} + L^2/(\mu_\mathrm{FE} C(V_\mathrm{g} - V_\mathrm{0}))
\end{equation*}

The fit is performed for $V_\mathrm{g}\geq V_\mathrm{infl}+2(V_\mathrm{infl}-V_\mathrm{th})$ with $R_\mathrm{S}$ and $\mu_\mathrm{FE}$ as fit parameters. This ensures an accurate estimate of $R_\mathrm S$ as confirmed on datasets with known $R_\mathrm S$ (see Supplementary Information Fig. \textbf{S7}).\\

\emph{\textbf{Step 5}: Use $V_\mathrm{0}$, $R_\mathrm{S}$ and $C$ to calculate $\mu_\mathrm{\textit{n}2m}(V_\mathrm{g})$}
\\

Using the value for $V_\mathrm{0}$ found in step 3, the $R_\mathrm{S}$ value found in step 4 and the measured $G(V_\mathrm{g})$ the gate-dependent $\mu$ is obtained from \begin{equation*}
\mu_\mathrm{2T}(V_\mathrm{g})=\frac{L^2}{C(V_\mathrm{g}-V_\mathrm{0})(G(V_\mathrm g)^{-1}-R_\mathrm{S})}
\end{equation*}
Together with $n(V_\mathrm g)$ obtained in step 3, this completes the procedure\\

\end{minipage}
\end{widetext}
\end{tcolorbox}
\end{figure*}

\begin{figure*}
\includegraphics[width=1\textwidth]{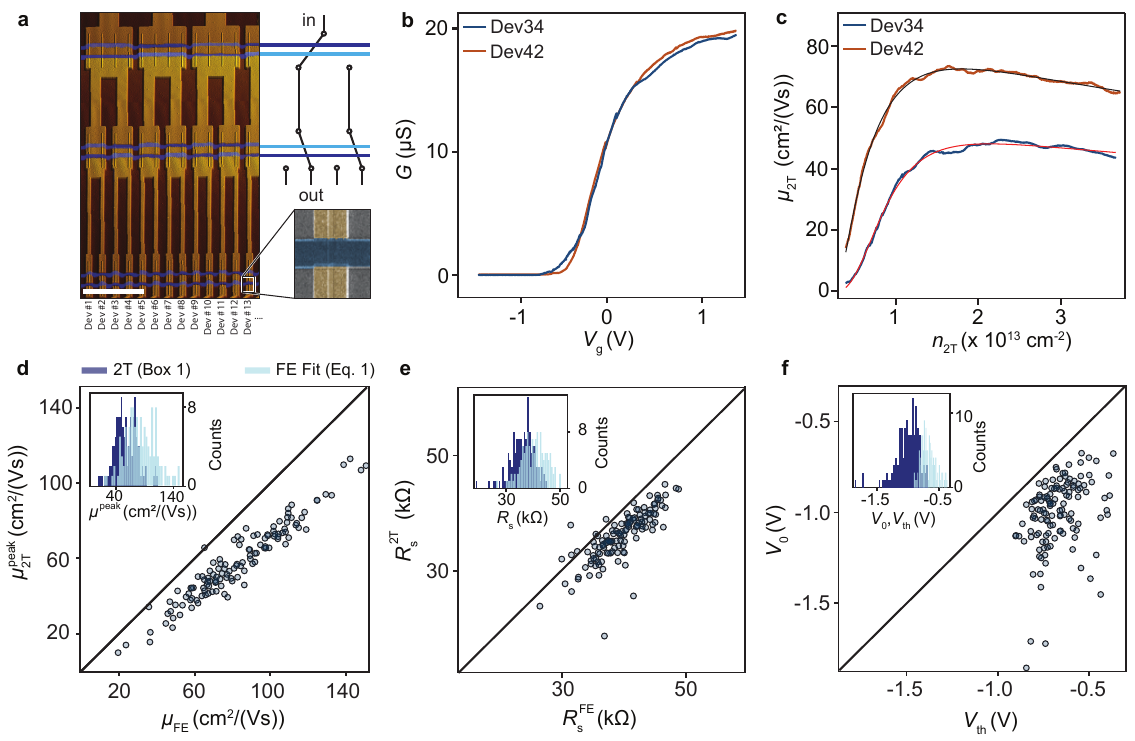}
\caption{\textbf{a}, Atomic force micrograph of a segment of the on-chip multiplexer circuit used to characterize a large number of devices~\cite{Ol_teins_2023}. Light and dark blue lines indicate the switching gates used to control the underlying nanowire FETs. In dark blue areas, gates run directly above the wires, separated by a layer of HfO$_\mathrm{x}$, enabling electrostatic coupling. In light blue areas, gates are screened from the wires by the gold contacts, such that each gate lead only influences every other multiplexer lead.
The inset shows an SEM image of an individual FET nanowire device.  
\textbf{b}, 2-terminal $G(V_\mathrm{g})$ for two representative nanowire FET devices measured at $T=$ 20 mK.  
\textbf{c}, Extracted $\mu_\mathrm{2T}$ vs. $n$, fitted to a Matthiessen sum  $\mu(n)^{-1} = \mu_1(n)^{-1} + \mu_2(n)^{-1}$, where $\mu_{1,2}(n) = A_{1,2}n^{\alpha_{1,2}}$.  
\textbf{d,e,f}, Comparison of parameters extracted by the new fitting routine (y-axis) with those obtained by the conventional fit model of Eq.~\ref{2point1} (x-axis). The analysis was performed on $G(V_\mathrm{g})$ data from 128 individual nanowire FETs. Insets show the comparison in the form of histograms.
}
{\label{fig:Fig2}. }
\end{figure*}
Measuring the device in a four-terminal configuration, including the transverse potential in an applied magnetic field, allows for conventional Hall analysis (Fig. \textbf{S3}). The Hall mobility, $\mu_\mathrm{Hall}(V_\mathrm g)$, and density $n_\mathrm{Hall}(V_\mathrm g)$ are shown in Fig.\ 1b. The density is linear in $V_\mathrm g$, and the slope yields a capacitance $C^i_\mathrm{Hall}=(6.2 \pm 0.3)\ \times 10^{-3}\ \mathrm{F/m^2}$ Which closely matches the results obtained with finite element modeling: $C^i_\mathrm{mod}=6.7\ \times 10^{-3}\ \mathrm{F/m^2}$. This confirms that the modeled capacitances can be used for devices where direct measurements of $n$ are not feasible.


In contrast to $\mu_\mathrm{FE}$, $\mu_\mathrm{Hall}$ shows a strong gate dependence (density) as expected from Hall characterization of 2D semiconductors\cite{ahn2021estimating,gazibegovic:2017}. At low values of $n$, $\mu_\mathrm{Hall}$ increases with $V_g$ consistent with transport being strongly affected by scattering from charged impurities, which are suppressed due to screening when increasing $n$\cite{Ahn_2022,makAPL2010,kawaharazukaJAP1999,SodergrenIEEE2023}. At higher values of $n$, $\mu_\mathrm{Hall}$ saturates and eventually decreases with $n$. This behavior is generally attributed to increased inter-subband scattering at high $n$ and/or surface and interface scattering with increased contribution as carriers accumulate towards the gate\cite{Sarma_2005,aseevBallisticInSbNanowires2019,Ahn_2022}.

The peak mobility $\sim5000\ \mathrm{cm^2/(Vs)}$ is observed at $V_\mathrm{g}\sim0.5$~V corresponding to a density of $n_\mathrm{Hall}\sim2\cdot10^{12}\mathrm{cm}^{-2}$. Notably, the $n$-independent $\mu_\mathrm{FE}$ exceeds $\mu_\mathrm{Hall}$ for all $V_\mathrm{g}$. From the $V_\mathrm{g}$-dependence of the densities shown in the inset, it is clear that this is related to an underestimation of $n$ by the FE-model which assumes a vanishing $n$ at $V_\mathrm{th}$, which is inconsistent with $G$ and $n_\mathrm{Hall}$ which both remain finite for $V_\mathrm{g} \leq V_\mathrm{th}$ (Fig.~1\textbf{b} inset). In turn, this leads to an overestimated $\mu$.

Using the Hall characterization as a benchmark, we developed a new procedure allowing the extraction of the, density, $n_\mathrm{2T}$, and $n_\mathrm{2T}$-dependent mobility, $\mu_\mathrm{2T}$, from the two-terminal $G(V_\mathrm{g})$ data. The resulting $\mu_\mathrm{2T}$ and $n_\mathrm{2T}$ are presented in Figure 1\textbf{b} and provide an excellent match to the Hall results. To improve the field-effect fitting based on Eq.\ 1, we rely on the following observations from the Hall characterization: Firstly, $n_\mathrm{Hall}(V_g)$ depends linearly on $V_g$ and can thus with high accuracy be described by a constant capacitor model, which disregards the energy-dependent density of states of the nanostructure. Secondly, the relevant capacitance can be accurately determined by standard finite-element calculations (see also Fig. \textbf{S2}) and therefore does not need to be treated as a fit parameter. Lastly, as discussed below, $n$ extrapolates to zero at $V_\mathrm{g}=V_0<V_\mathrm{th}$. With a procedure for estimating $V_0$ a quantitative account of $\mu(n)$ can be obtained by modifying Eq.\ \ref{2point1} to account for a $n_\mathrm{2T}$-dependent $\mu_\mathrm{2T}$,
\begin{equation}
\mu_\mathrm{2T}(V_\mathrm{g})=\frac{L^2}{C(V_\mathrm{g}-V_\mathrm{0})(G(V_\mathrm g)^{-1}-R_\mathrm{S})}
\end{equation}
To estimate $V_0$, we note that at low $n$, transport is strongly influenced by the random potential from charged impurities in the vicinity of the channel as schematically shown in the inset to Fig.\ 1\textbf{a}. Upon increasing $V_g$ (Fermi energy, $E_\mathrm F$), the electrons that initially occupy the conduction band at $V_\mathrm{0}$ appear in spatially disconnected regions, while the onset of electronic transport occurs at \ $E_\mathrm F \sim E_\mathrm{th}$ at  the percolation threshold ($V_\mathrm{g} \sim V_\mathrm{th}$) near half-max of the potential fluctuation landscape~\cite{aharony2003percolation,Parviainen_2007}. Upon further increasing $V_\mathrm g$, a crossover to a regime described by conventional Drude scattering occurs for $E_\mathrm F$ near the top of the potential fluctuation landscape\cite{NixonPRB1990,TracyPRB09}. As shown in Supplementary Fig.\ \textbf{S4} this transition occurs at the inflection point of $G(V_\mathrm{g})$, $V_\mathrm g \sim V_\mathrm{infl}$. The difference $V_\mathrm{th}-V_\mathrm{infl}$ sets a characteristic scale of the potential fluctuation amplitude and thus $V_\mathrm{0}=V_\mathrm{th}-m(V_\mathrm{infl}-V_\mathrm{th})$ for some factor $m$. Compared with the Hall measurements, we empirically find $m=2$ for our devices (see Fig. \textbf{S5}). Although the exact value of \( m \) depends on the details of the potential landscape, the limits \( m \ll 1 \) and \( m \gg 1 \) correspond to unphysical scenarios: the first implies \( V_0 = V_\mathrm{th} \), and the latter implies an abrupt onset of Drude transport (as \(  V_\mathrm{infl} \rightarrow V_\mathrm{th} \)). These considerations constrain physically meaningful values of \( m \), even when the percolation threshold deviates from the half-max of the potential landscape. Arguably, since \( V_\mathrm{infl} \) must fall between \( V_\mathrm{th} \) and the top of the potential, assuming that $V_\mathrm{th}$ corresponds to $E_\mathrm{F}$ halfway is a reasonable first approximation. This assumption yields a value of \( V_0 \) that supports our empirical choice of \( m = 2 \) as a physically motivated approximation.  An additional discussion of the potential limits on \( m \) is provided in Fig.\ \textbf{S5}. 
Furthermore, as shown in supplementary Fig.\ \textbf{S5}, the shape of the $\mu_\mathrm{2T}$ curves remains invariant for \( m \geq 0.4\), demonstrating the robustness of the method. 

The final parameter in the fit is $R_\mathrm{S}$, estimated by noting that it dominates the conductance at high $V_g$ and can be accurately found by fitting Eq.\ \ref{2point1} in the restricted range $V_\mathrm{g} \gg V_\mathrm{infl}$, and replace $V_\mathrm{th}$ by $V_\mathrm{0}$.

%

%

In Box 1, we provide a step-by-step guide for implementing these steps illustrated by application to the case of the two-terminal measurement of $G(V_\mathrm g)$ from Fig.\ 1\textbf a. The resulting $\mu_\mathrm{2T}(V_\mathrm g)$ and $n_\mathrm{2T}(V_\mathrm g)$ are shown in Fig.~1b and are in excellent agreement with the results of full Hall characterization of the same device. Three additional Hall devices were analyzed and compared in the same way, and all show the same excellent consistency (Supplementary Fig.\ \textbf{S6}). The slight difference can be attributed to the differences in the regions of the device probed by the two measurement schemes: the four-terminal Hall measurement characterizes the nanowire segment between the longitudinal voltage probes, while the two-terminal measurement probes the entire length below the gate. Indeed, as a confirmation, an even better match of $\mu_\mathrm{2T}$ and $\mu_\mathrm{Hall}$ is obtained when extracting $\mu_\mathrm{2T}$ using the four-terminal conductance (Supplementary Fig. \textbf{S7}). Thus, Fig.\ 1 confirms that by using the procedure in Box~1, the density dependent mobility can be reliably extracted, providing the same information as multi-terminal Hall measurements, however, relying on the much more accessible two-terminal FET measurements. This is the central result of this manuscript.

\section*{Statistical characterisation and reproducibility}
Since the majority of existing literature uses $\mu_\mathrm{FE}$ (Eq.\ \ref{2point1}) as a measure of the mobility, we now systematically compare the parameters obtained by the FE analysis and our new improved method. To this end, measurements of 128 nominally identical nanowire FET devices were characterized by both methods. The FETs are based on single-axis nanowires where Hall measurements, such as in Fig.\ 1, are unfeasible \cite{storm:2012,blomers:2012,gluschkeImpactInvasiveMetal2020a,choiAFM2018}. The devices are based on arrays of selective area grown InAs nanowires with a structure similar to the Hall bar in Fig.\ 1, except for the omission of the InGaAs buffer layer. Ohmic Ti/Au contacts were defined using e-beam lithography directly on the growth substrates with a channel length of $1 \, \mu \mathrm m$ and a top-gate covered the entire channel. Figure 2\textbf{a} shows an example of a device. Individual FET devices were measured using an on-chip multiplexer/demultiplexer circuit as discussed in detail in Ref.~\citenum{Ol_teins_2023}.  Figure~2\textbf{a} shows a section of the 8-level multiplexer circuit, which can address any one of 256 devices. Figure 2 shows results from analysis of 128 devices measured at $T=20$~mK, and the corresponding results from all 256 devices at $T=100$~K are presented in Supplementary Fig.\ \textbf{S8}$\&$\textbf{S9}.

Figure 2b shows the transfer characteristics $G(V_\mathrm g)$ for two representative devices, and Fig.\ 2c shows the corresponding $\mu_\mathrm{2T}(n_\mathrm{2T})$ extracted following the steps discussed above (Box 1). For both devices, $\mu_\mathrm{2T}$ is qualitatively similar to those of the Fig~1. Figure~2\textbf{d-f} show the correlations between parameters extracted by the conventional field-effect model (Eq. \ref{2point1}) and the method presented here (Box 1). Inset histograms show the frequency of each parameter value across the ensemble. For $\mu$, the FE-model yields only a single parameter while our extended procedure provides the relationship $\mu_\mathrm{2T}(n_\mathrm{2T})$. To compare, the peak value, $\mu_\mathrm{2T}^\mathrm{peak}$, is extracted and plotted in Fig.\ 2d against $\mu_{FE}$. Clearly, the conventional FE model overestimates even the peak mobility by about $75\%$, and for all devices $\mu_\mathrm{2T}^\mathrm{peak}<\mu_\mathrm{FE}$. This is a consequence of the finite density at $V_\mathrm{th}$, which is not taken into account by the FE model and thus $V_\mathrm{0}<V_\mathrm{th}$ (Fig.~2\textbf{f}). We note that this factor could adjust previous $\mu$-estimates based on Eq.\ 1 to give a more realistic value. For the same reason and the restricted fitting range (see Box 1) the new procedure yields a lower value for $R_\mathrm{S}$ (Fig.\ 2e). 


Finally, we note that the spread of the extracted parameters in the histograms in Fig.\ 2d-f is similar for both methods, which shows that the uncertainty is not compromised, although the entire $n$-dependency is obtained.

\begin{figure}
\includegraphics[width=0.45\textwidth]{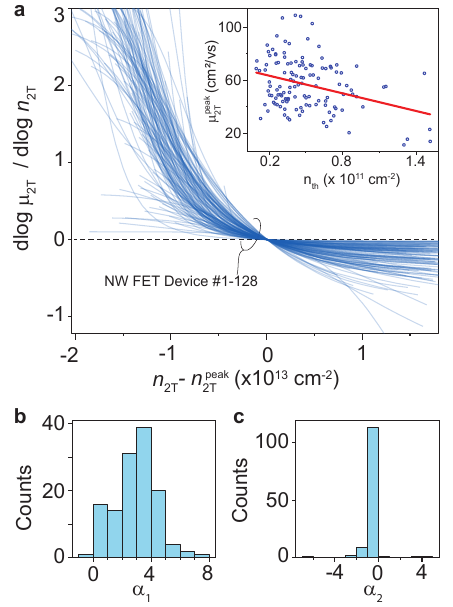}
\caption{\textbf{a.} Mobility exponent $\alpha(n_\mathrm{2T})$ of nanowire devices $\# 1-128$ defined as the log derivative of $\mu_\mathrm{2T}$. Inset shows the peak mobility vs.\ the density at the threshold. \textbf{b,c} histograms of the two exponents obtained from fitting all $\mu_\mathrm{2T}$ curves to a Matthiessen sum $\mu_\mathrm{2T}^{-1} = \mu_1(n)^{-1}+\mu_2(n)^{-1}$.\label{fig:Fig35} }
\end{figure}

\section*{Analysis of scattering mechanisms}
The non-monotonic shape of $\mu_\mathrm{2T}$ in Fig. 2c reflects the $n$-dependence of different scattering mechanisms that limit $\mu$ in nanowires, as previously discussed\cite{DasSarmaPRB2013,ahn:2021}. Upon increasing $n$, an initial steep rise of $G$ and $\mu_\mathrm{2T}$ is observed at the percolation threshold $V_\mathrm{th}$, followed by an intermediate regime, where long-range Coulomb scattering by charged impurities in and around the conducting channel is the dominant scattering mechanism. The scattering becomes less effective with increasing $n$ due to screening, leading to an increasing $\mu_\mathrm{2T}$ toward its peak value. With further increasing $n$, surface and inter-subband scattering become increasingly stronger, leading to the observed saturation and eventual drop in $\mu_\mathrm{2T}$ at high $n$. In Refs.\ \citenum{DasSarmaPRB2013,ahn:2021}, the scattering problem was treated by Boltzmann transport theory, predicting a universal $\mu$ scaling $\mu(n) \propto n^{3/2}$ with an exponent of $3/2$ in the intermediate $n$ regime. In Fig.\ 3a we show the experimental $\mu_\mathrm{2T}$ exponent $\alpha(n_\mathrm{2T}) = \mathrm d \log \mu_\mathrm{2T} / \mathrm d \log n_\mathrm{2T}$ extracted as the derivative of the log-transformed $\mu_\mathrm{2T}$ for all the nanowire devices. To allow easier comparison between devices, $\alpha(n_\mathrm{2T})$ is shown for $n_\mathrm{2T}-n_\mathrm{2T}^\mathrm{peak}$ where $n_\mathrm{2T}^\mathrm{peak}$ is the value of $n_\mathrm{2T}$ corresponding to $\mu_\mathrm{2T}^\mathrm{peak}$. The $\mu_\mathrm{2T}$ exponent peaks at low $n_\mathrm{2T}$ towards the percolation threshold as expected and drops monotonously towards $n_\mathrm{2T}^\mathrm{peak}$. The absence of pronounced plateaus shows that the nanowires are not strongly dominated by a single scattering mechanism at any $n_\mathrm{2T}$, a result which is consistent with the analysis in Ref.\ \citenum{ahn:2021}. In this scenario, the log-derivative, $\alpha(n_\mathrm{2T})$, can attain non-universal values even though each of the underlying scattering mechanisms may be characterized individually by universal exponents, making a direct comparison between theory and experiment difficult. As a first approach, however, we show in Fig.\ 2c fits of $\mu_\mathrm{2T}$ to a simple two-term Matthiessen sum $\mu_\mathrm{2T}^{-1} = \mu_1(n)^{-1}+\mu_2(n)^{-1}$ where $\mu_{1,2}(n) = A_{1,2}n^{\alpha_{1,2}}$. This provides an excellent description of $\mu_\mathrm{2T}$ in the measured range, and Fig.\ 3b,c show histograms of the two exponents for fits to all of the 128 nanowire devices. The increase in $\mu_\mathrm{2T}$ in the intermediate $n_\mathrm{2T}$ regime is characterized by an average value of $\alpha_1 \approx 3$, which exceeds the theoretical expectation for charged impurity scattering by a factor of two, which we attribute to the fitting range extending into the percolation regime, where $\alpha(n_\mathrm{2T})$ diverges, thereby inflating $\alpha_1$. As a result, the extracted value of $\alpha_1$ is sensitive to the choice of fitting range.
We note that in this regime, the impurity concentration, which is on the order of the percolation threshold $n_\mathrm{th}$, should constitute the limiting factor for $\mu_\mathrm{2T}^\mathrm{peak}$\cite{ahn:2021}. Accordingly, the inset to Fig.\ 3a shows $\mu_\mathrm{2T}^\mathrm{peak}$ vs.\ $n_\mathrm{th}$ for all 128 nanowire devices. A weak decreasing trend is observed, consistent with this scenario; however, a significant spread is present and shows that other effects are dominating the obtained $\mu_\mathrm{2T}^\mathrm{peak}$. Further experiments studying the effects of varying nanowire growth conditions, dimensions, dielectric environments, etc.\, and exploiting the statistical analysis of $n$-dependent $\mu$, enabled by the method described here, should enable a much-needed insight into the scattering mechanisms and provide a strategy for improving $\mu$.

\begin{figure}
\includegraphics[width=0.45\textwidth]{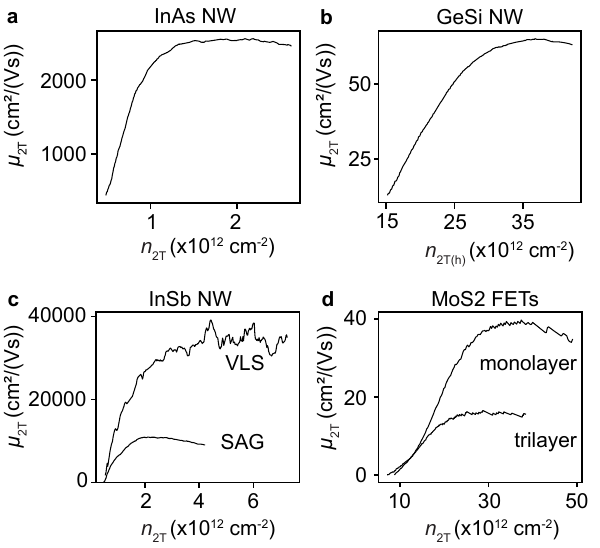}
\caption{Effective mobilities, $\mu_\mathrm{2T}$ extracted from $G$ vs. $V_\mathrm{G}$ curves of FETs that were measured by other groups. We used a web digitizer tool to retrieve the data sets from the figures in the papers. 
\textbf{a}, is from Ref. \cite{ford:2009}. \textbf{b}, is from Ref. \cite{GeSi}. \textbf{c}, is from \cite{aseevBallisticInSbNanowires2019} for SAG and Ref. \cite{gulHighMobilityInSb2015} for VLS . \textbf{d}, is from Ref.\cite{Krasnozhon}}
\end{figure}
\section*{Re-analysis of existing results}
Finally, to emphasize the significance and usefulness of the new method, we demonstrate the general applicability of the procedure to other types of nanostructures and materials systems beyond the InAs SAG nanowires discussed above. For a given nanostructure the method requires only measurements of the two terminal FET-type transfer characteristics $G(V_\mathrm{g})$ and relevant geometrical parameters and thus enables re-analysis of existing data previously only treated by the FE model of Eq. 1. As examples, Fig.\ 4 shows $\mu_\mathrm{2T}$ extracted for six different device classes from the existing literature: $n$-type VLS grown InAs nanowires\cite{ford:2009},  $p$-type, VLS grown SiGe nanowires\cite{GeSi}, $n$-type InSb nanowires for both SAG and VLS growth\cite{aseevBallisticInSbNanowires2019, gulHighMobilityInSb2015}, and FET devices based on 2D monolayer and trilayer MoS$_2$\cite{Krasnozhon}. In all cases, the $n$-dependent $\mu_\mathrm{2T}$ could be reliably extracted, thus providing new insights into the transport characteristics of these systems. A detailed discussion of the significance of $\mu(n)$ in each case is outside the scope of the present work; however, the results illustrate the general potential of our procedure.

\section*{Discussion}
The presented method for extracting $\mu_\mathrm{2T}(n)$ enables analysis of scattering mechanisms in nanomaterials where fabricating more than two contacts is unfeasible and/or external magnetic fields are unavailable. The method assumes transport in a 2DEG at the semiconductor/insulator interface and the existence of a charged impurity-generated background potential landscape, which dominates transport at low density and causes a classical metal-insulator transition between Drude and percolation transport. These requirements hold for devices based on, among others, III-V materials, van der Waals materials, and group-IV materials. Given the standard deviation of the potential fluctuations is expected to be on the order of 20~meV even for high-$\mu$ GaAs/AlGaAs heterostructures\cite{NixonPRB1990}, the method should be valid up to room temperature for most materials. However, alternative methods for finding $V_\mathrm{0}$ may need to be developed for device materials where different physics dominates at low density. Additionally, the estimation of $R_\mathrm{S}$ relies on a weak dependence of $\mu$ on $n$ at high $V_\mathrm{g}$, requiring both that this holds true for the sample in question and that data is acquired to sufficiently high $V_\mathrm{g}$. Note, however, that increasing $V_g$ range may introduce sweep-rate dependent hysteresis due to filling/emptying of charge traps. Finally, throughout this manuscript, we have ignored the method for estimating $\mu(V_\mathrm{g})$ by taking the derivative $dG/dV_\mathrm{g}$ of the Drude expression\cite{doi:https://doi.org/10.1002/0471749095.ch8}. This method is not only problematic because it ignores $R_\mathrm{S}$, but also assumes a $V_\mathrm{g}$ dependent $\mu$ while not invoking the chain rule in performing the derivative. This method is therefore mathematically invalid, and we show in Supplementary Section 2 that properly employing the chain rule results in our expression for $\mu_\mathrm{2T}$ used throughout.

\section{Methods}

The devices were based on SAG nanostructures and wires\cite{OriginalSAG} using procedures described in refs.~\citenum{beznasyukPRM22} and \citenum{Ol_teins_2023}. The resultant structures feature a two dimensional electron gas at the surface. Ohmic contacts consisted of a 5/250 nm Ti/Au stack and were fabricated using electron beam lithography (EBL) and e-beam evaporation. Ar ion milling was used to remove native oxide from the nanowire immediately prior to deposition. A 15 nm HfO$_x$ dielectric layer was deposited using atomic layer deposition, and the 5/250 nm thick Ti/Au top gate was defined again using EBL and e-beam evaporation.

Electron transport measurements were conducted in a dilution refrigerator with base temperature $\sim 20 \, \mathrm{mK}$ and equipped with a 6-1-1 T vector magnet. The currents and voltage drop was measured using standard audio-range lock-in techniques. Low-pass filtering of measurement lines at $\sim 20 \, \mathrm{mK}$ adds $\sim 7.2 \, \mathrm k \Omega$ in series to the device. In both configurations, voltage $V$ was applied to one of the end contacts, and the resultant current $I$ measured at the other, with $V_\mathrm{g}$ used to alter $n$. To conduct a Hall measurement, longitudinal and transverse voltages, $V_\mathrm{xx}$ and $V_\mathrm{xy}$ were measured in response to a magnetic field $B$ applied perpendicular to the substrate plane, with the fourth voltage probe floating. To conduct two-terminal FET measurements, all voltage probes were floated, and only the two-terminal conductance was measured.

\begin{acknowledgments}
This research was supported by research grants from European Research Council under the European Union’s Horizon 2020 research and innovation program (Grant nos. 716655 and 866158) (T.S.J.). The Authors acknowledge useful discussions with Peter Krogstrup.
\end{acknowledgments}

\bibliography{apssamp,TSJrefs}

\end{document}


\maketitle
\clearpage
\section*{Section 1 - Supplementary figures}
\subsubsection*{S1. Details of device fabrication, electrical circuit setup and composition of the Hall bar nanotructures}

\textbf{S1} Figure \textbf{S\ref{fig:suppFig1}} shows an example of a Hall-bar device that was fabricated using procedures described in Ref. \cite{beznasyuk:2021} with the 4-side arms SAG NWs covered by ohmic contacts (yellow leads). The top gate (orange) was deposited to control the carrier density in the devices. Both the ohmic contacts and the gate are constructed with 5 nm thick Ti and 250 nm thick Au, and the gate was isolated from the NW and the ohmic contacts by a 15 nm thick HfOlayer $_2$, which is deposited by the deposition of the atomic layer. The inset shows an illustration of the cross-section of the wire. Electron transport measurements were performed in a dilution refrigerator with a base temperature $\sim 20 \, \mathrm{mK}$ and equipped with a 6-1-1 T vector magnet. Currents and voltage drops were measured using standard lock-in techniques. The low-pass filtering of the measurement lines at $\sim 20 \, \mathrm{mK}$ adds $\sim 7.2 \, \mathrm k \Omega$ in series to the device. In both the FET and the Hall configuration, a voltage $V_\mathrm{AC}$ was applied to one of the end contacts, and the resulting current $I$ was measured vs. the top-gate voltage $V_\mathrm{g}$ at the other. To conduct a Hall measurement, longitudinal and transverse voltages, $V_\mathrm{xx}$ and $V_\mathrm{xy}$ were measured in response to a magnetic field $B$ applied perpendicular to the substrate plane, with the fourth voltage probe floating. To conduct two-terminal FET measurements, all voltage probes were floated and only the two-terminal conductance was measured.

\begin{suppfigure}[h]
\includegraphics[width=1.0\linewidth]{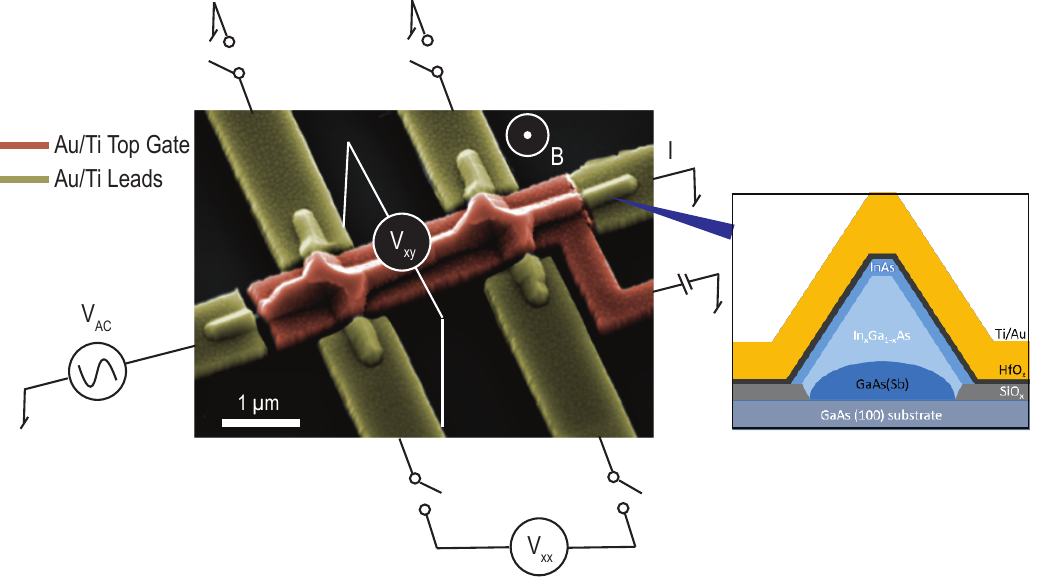}
\caption{False-colored SEM of the finished Hall bar with a schematic of the circuit used in the electrical characterization. The inset illustrates a cross-section of the wire. .}
\label{fig:suppFig1} 
\end{suppfigure}
\clearpage

\subsubsection*{S2. Finite element simulations}

3D COMSOL finite element simulations were used to simulate both the Hall resistance and capacitance of the nanowires. In the model, the relation between the current, $\textbf{j}$, and the electric field, $\textbf{E}$, in the presence of a magnetic field, $\textbf{B}$, is given by the expression:

\begin{equation*}
   \textbf{j} = \sigma(\textbf{B}) \cdot \textbf{E}
   \label{Eq.Ohm}
\end{equation*}

\noindent where $\sigma(B)$ is given by:

\begin{equation*}
    \sigma(B) = \frac{\sigma_{0}}{1+\beta^2} \begin{bmatrix} 1 &  -\beta &  0 \\ \beta & 1 & 0 \\ 0 & 0 & 1 \end{bmatrix}
\end{equation*}

\noindent $\sigma_0$ is the conductivity of the material in the absence of a magnetic field:

\begin{equation*}
   \sigma_0 = ne\mu
\end{equation*}

\noindent $n$ is the carrier density, $\mu$ is the carrier mobility, and $\beta$ is the unitless magnetic field:

\begin{equation*}
   \beta = \mu B
\end{equation*}

The nanowire geometry used in the simulations was made to mimic that of the experimental devices. The total length of the hall cross was set to 6 $\mu$m with an arm spacing of 2 $\mu$m. The nanowires feature a isosceles triangular cross-section where the bottom face is 215 nm across and forms a 55$^\circ$ angle with the other two faces. InAs nanowires conduct via a 2DEG layer located at the surface. To simulate the conduction through a 2DEG, a core-shell model was used with a conducting InAs surface layer and non-conducting InGaAs core. A volume was defined which consisted of all the material located within 20 nm of the top two faces of the triangular cross-section in order to represent the 2DEG. The carrier density, $n$, and electron mobility, $\mu$, in this region are gate dependent and were determined via experiments. Polynomial fits of the data were implemented in COMSOL to set the material properties for a given gate voltage, $V_g$:

\begin{equation*}
    \mu = -9521.7 V_g^4 + 37582 V_g^3 - 54425 V_g^2 + 33023 V_g -1190.5  \hspace{0.1cm} (cm^2/Vs)
\end{equation*}

\begin{equation*}
    n = 4.54 \cdot 10^{12} \hspace{0.1cm} V_g + 3.40 \cdot 10^{11} \hspace{0.1cm} (cm^{-2})
\end{equation*}

A HfO$_2$ dielectric layer extending 10 nm from the surface of the nanowire structure was also defined followed by a volume of Au to simulate the top gate. The geometry of the simulate device is can be seen in Fig. \textbf{S\ref{fig:suppFig2}c} along with a cross-section of the nanowire structure in Fig. \textbf{S\ref{fig:suppFig2}a}. The actual experimental devices featured a misalignment in the pair of arms used as transverse voltage probes which resulted in an offset of the Hall data due to intermixing of the $R_\mathrm{xx}$ and $R_\mathrm{xy}$ components. Analysis of electron microscopy images revealed that the misalignment was approximately 12.5\% of the distance between the longitudinal probes. This misalignment was also implemented in the COMSOL model. Both the experimental and simulated had 0.125$R_\mathrm{xx}$ subtracted from the $R_\mathrm{xy}$ signal in order to compensate for the $R_\mathrm{xx}$ contribution. It was observed that adjusting the experimentally determined carrier density by a factor of 0.9 produced the closest fit to the data as shown in Fig. \textbf{S\ref{fig:suppFig2}d}. The small asymmetry in the $R_\mathrm{xy}$ signal about 0 T which is captured by the simulation is likely a result of the geometric cross-section of the structure. 

Additionally, the capacitance of the device was also simulated and yielded a value of $1.375 \cdot 10^{-14}$ F, corresponding to an capacitance per unit area of $6.683 \cdot 10^{-3}$ F/m$^2$, which is in excellent agreement with the experimental results. The distribution of the internal electrical field upon the application of a gate voltage can be seen in Fig. \textbf{S\ref{fig:suppFig2}b}.

\begin{suppfigure}[h]
\includegraphics{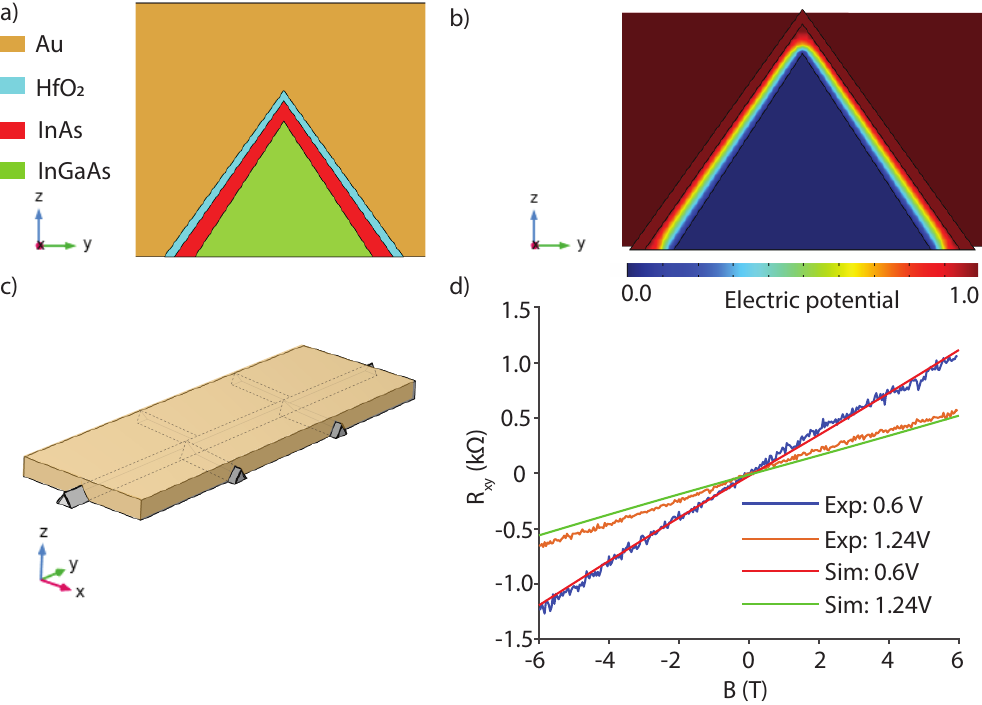}
\caption{ \textbf{a} Schematic of the cross-section of the modeled device with the different material regions indicated. \textbf{b} Distribution of the internal electrical potential upon the application of a voltage to the top gate. \textbf{c} Overall geometry used in the capacitance simulations. \textbf{d} Experimental Hall resistance of a nanowire structure at two different gate voltages plotted alongside the simulated results. \label{fig:suppFig2} }
\end{suppfigure}

\subsubsection*{S3. Electrical characterization of the Hall device}
Figure \textbf{S\ref{fig:suppFig3}a} shows the raw data from the electrical measurements of the Hall bar conducted in a 4-terminal ($R_\mathrm{xy}$ in the left plot and $R_\mathrm{xx}=1/G_4$ in the middle plot) and 2-terminal (right plot) configuration vs. $V_\mathrm{g}$ and $B_\perp$. Line cuts of the 2d colorlpots are shown in \textbf{S\ref{fig:suppFig3}b$\&$c}. The opaque lines in \textbf{S\ref{fig:suppFig3}b} shows the raw data while for the less opaque lines a component of 0.125$R_\mathrm{xx}$ has been subtracted from $R_\mathrm{xy}$. The latter were used to extract the Hall density for each line cut along $V_\mathrm{g}$. \par\medskip

\begin{suppfigure}[h]
\centering
\includegraphics[width=1\linewidth]{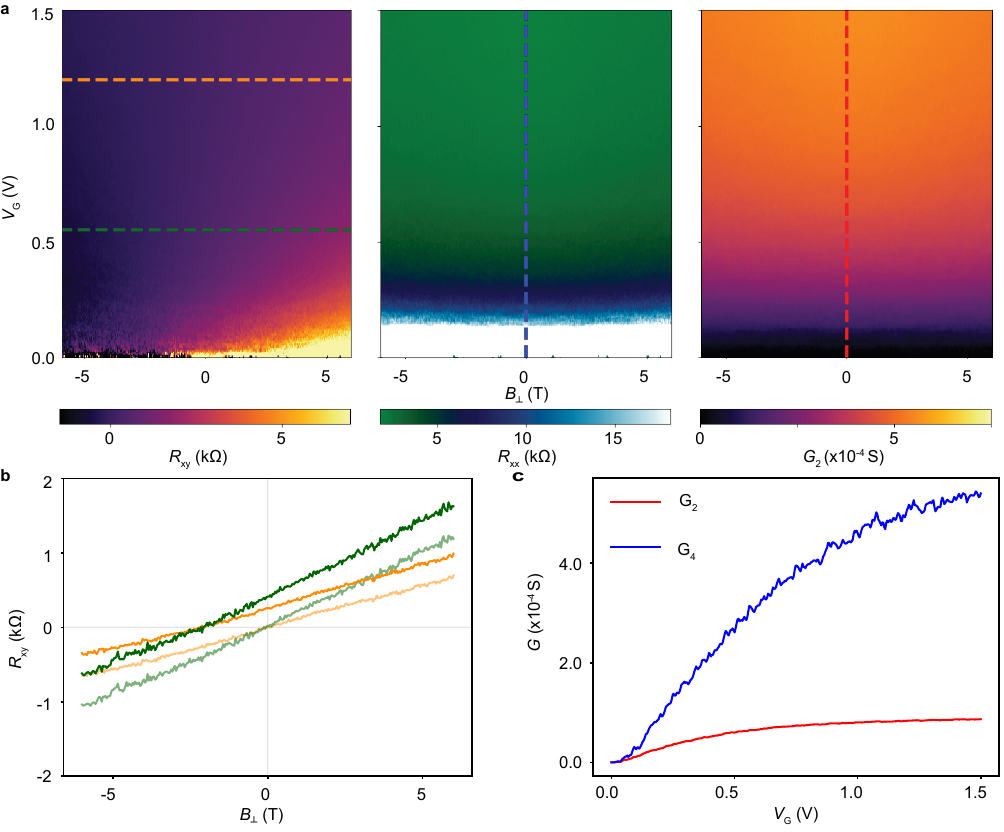}
\caption{ In \textbf{a}, we show the Raw data of Hall bar. \textbf{\textit{Left:}} $R_\mathrm{xy}$ vs. $V_\mathrm{g}$ and $B_\perp$. \textbf{\textit{Middle:}}   $R_\mathrm{xx}$ vs. $V_\mathrm{g}$ and $B_\perp$. \textbf{\textit{Right}}   $G_\mathrm{2}$ vs. $V_\mathrm{g}$ and $B_\perp$. In \textbf{b}, the line traces of the left 2D-plot in \textbf{a} are shown with opaque curves showing the raw data and the less opaque curves  shows $R_\mathrm{xy}$ after a $0.125· R_\mathrm{xx}$ component has been subtracted.  In \textbf{c}, we show line traces of the middle 2D-plot ($G_\mathrm{4}$) and right 2D-plot ($G_\mathrm{2}$) in \textbf{a}, comparing the 2-terminal and 4-terminal measured conductance. \label{fig:suppFig3} }
\end{suppfigure}

\subsubsection*{S4. Metal-insulator transition at the inflection point} 

In the main text, we associate the inflection point of $G(V_\mathrm{g})$ traces with a cross-over between percolation and Drude transport. The percolation-to-Drude crossover has been extensively studied through the lens of a metal-to-insulator transition, since transport in the Drude regime has a markedly different temperature dependent behaviour (metallic, where $R$ increases with increasing $T$) compared to transport in the percolation regime (insulator-like, $R$ decreases with increasing $T$).\cite{TracyPRB09} The left-hand set of panels in Figure \textbf{S\ref{fig:suppFig4}} shows temperature-dependent $G(V_\mathrm{g})$ data from four devices within the multiplexer used in main text Fig.~2. At high(low) $V_\mathrm{g}$, we observe a decrease(increase) in conductance, consistent with a percolation-induced metal-to-insulator transition.

To examine the transition more closely, the right-hand set of panels shows the same data transformed to $R-R_\mathrm{C}$ vs $T$ where $R_\mathrm{C}$ is extracted as described in the main text Box. At low $V_\mathrm{g}$ (more purple-colored traces), the resistance decreases with increasing temperature (insulator-like), while traces at higher $V_\mathrm{g}$ (more yellow-colored traces) have an increasing $R$ with increasing $T$. Plotted in black is the extracted inflection point $V_\mathrm{infl}$ for each $T$. The inflection point remains relatively temperature-independent, and matches the $V_\mathrm{g}$ for which transport crosses over from percolation to Drude-like. We therefore associate $V_\mathrm{infl}$ with the voltage/density at which the devices undergo a transition between percolation and Drude-like transport.

\begin{suppfigure}[h]
\includegraphics[width=1.0\linewidth]{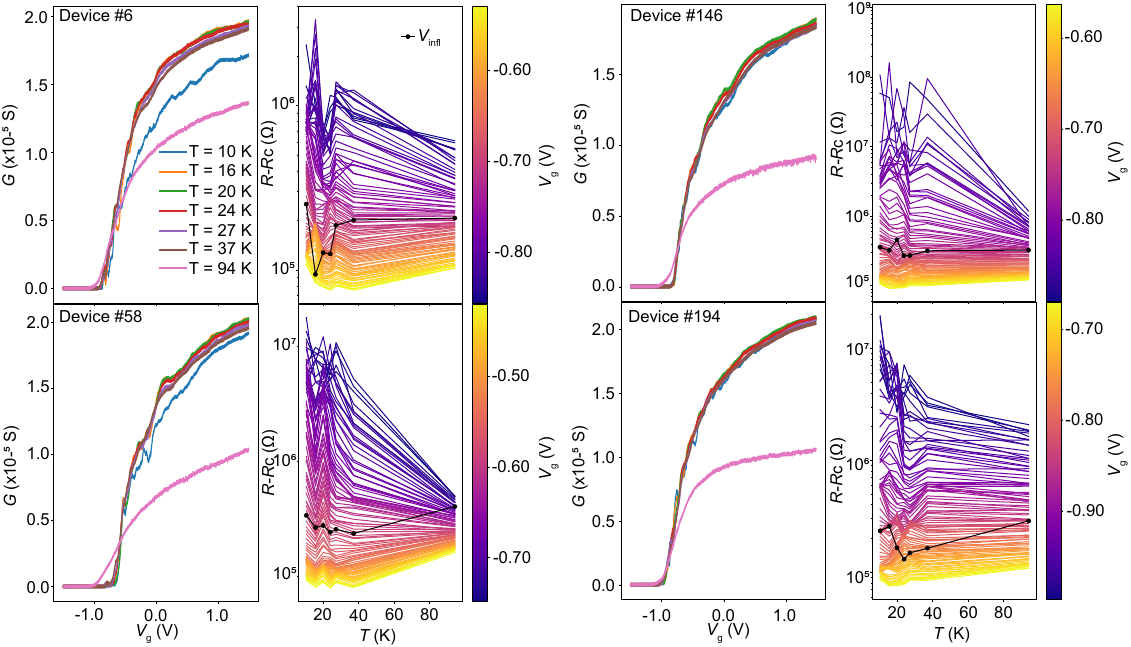}
\caption{$G_2$ vs. $V_\mathrm{g}$ and $R-R_\mathrm{c}$ vs. $T$ for 4 sets of SAG-based transistors from the multiplexer.\label{fig:suppFig4} }
\end{suppfigure}
\subsubsection*{S5. Empirical Analysis of the Factor $m$}

In Fig.\textbf{S\ref{fig:suppFig5}}, we compare the extracted $\mu_\mathrm{2T}$ for various values of the parameter $m$ using our method against the Hall mobility, $\mu_\mathrm{Hall}$ (yellow curve). We find that choosing $m = 2.0$ provides a good approximation (blue curve). Importantly, varying $m$ from 0.6 to 4.0 preserves the overall trend of $\mu_\mathrm{2T}$, demonstrating the robustness of the method. This trend invariance arises from the compensating behavior of the parameters $R_\mathrm{s}$ and $V_0$: as $m$ increases, $V_0$ decreases approximately linearly, while $R_\mathrm{s}$ also decreases proportionally. This mutual compensation effectively rescales the mobility without altering its shape, reinforcing the method's suitability for analyzing scattering mechanisms, where the rigidity of the curve shape ensures that extracted scattering parameters (i.e., $\alpha$ values) remain consistent.

The curves highlighted in purple, green, and red for $m = 0.0$, $0.2$, and $0.4$, respectively, diverge as $V_\mathrm{g}$ approaches zero from the right. This occurs because the model predicts a finite conductance at $V_0$, where the carrier density is defined to be zero  - an unphysical result. Therefore, we can confidently set a lower bound of $m > 0.4$.
\begin{suppfigure}[h]
\includegraphics[width=0.99\linewidth]{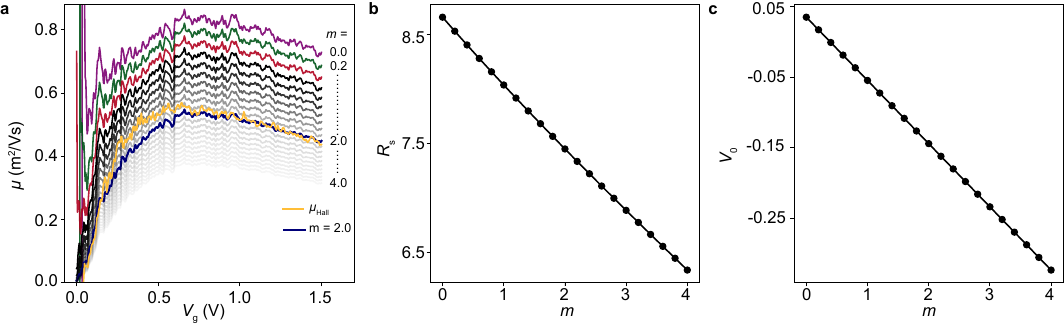}
 \caption{\textbf{a}, $\mu_\mathrm{2T}$ vs. $V_\mathrm{g}$ for various values of $m$ compared to the Hall mobility. \textbf{b}, $R_\mathrm{s}$ vs. $m$ . \textbf{c}, $V_\mathrm{0}$ vs. $m$\label{fig:suppFig5} }
\end{suppfigure}

\subsubsection*{S6. Meausurements of additional SAG Hall Bars}
Figure \textbf{S\ref{fig:suppFig6}} presents a comparison of $\mu_\mathrm{2T}$ and $\mu_\mathrm{Hall}$ from three additional Hall bars, each fabricated similarly to the device shown in \textbf{S\ref{fig:suppFig1}}. All devices exhibit excellent agreement between the two mobility measurements, demonstrating high consistency across samples.
\begin{suppfigure}[h]
\includegraphics[width=0.95\linewidth]{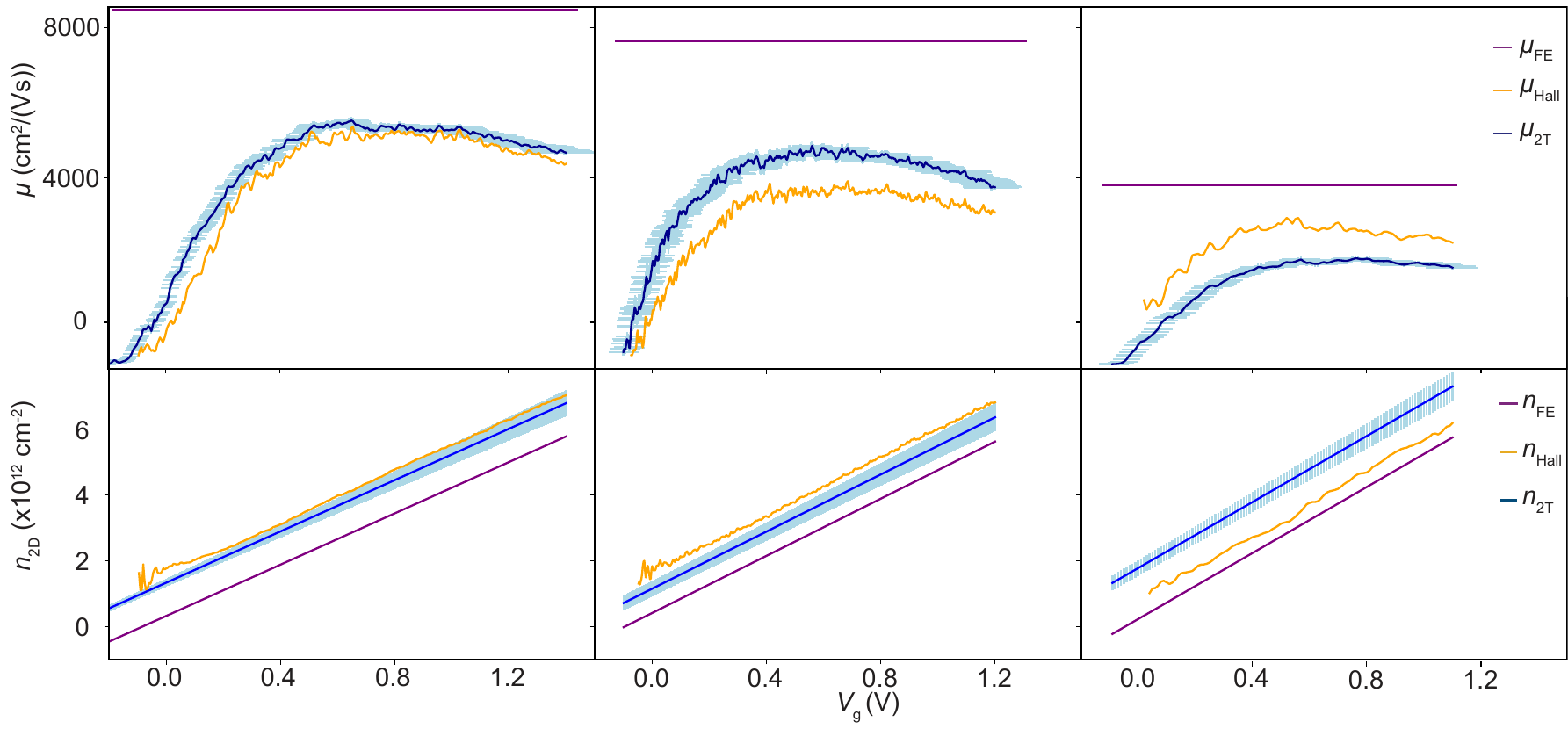}
\caption{$\mu_\mathrm{2T}$, $\mu_\mathrm{FE}$  and $\mu_\mathrm{Hall}$ vs. $V_\mathrm{g}$ for three additional Hall bars.\label{fig:suppFig6} }
\end{suppfigure}
\subsubsection*{S7. $\mu_\mathrm{2T}$ extracted using four terminal $G(V_\mathrm{g})$ compared to $\mu_\mathrm{Hall}$ and $R_\mathrm{s}$ fit analysis}
Figure \textbf{S\ref{fig:suppFig7}}a compares the mobility extracted from a 4-terminal measurement, $\mu_\mathrm{2T}^{G_4}$, with $\mu_\mathrm{Hall}$. As expected, $\mu_\mathrm{2T}^{G_4}$ shows close agreement with $\mu_\mathrm{Hall}$. In principle, $\mu_\mathrm{2T}^{G_4}$ and $\mu_\mathrm{Hall}$ should match exactly if $V_\mathrm{0}$ is perfectly estimated. The small discrepancy in the fluctuation amplitude between $\mu_\mathrm{2T}^{G_4}$ and $\mu_\mathrm{Hall}$ arises from the use of a estimated carrier density $n_\mathrm{2T}$ in the former versus a measured $n_\mathrm{Hall}$ in the latter.  The 2-terminal measured, $\mu_\mathrm{2T}^{G_2}$, and the standard $\mu_\mathrm{FE}$ are also shown in the plot.  Figure \textbf{S\ref{fig:suppFig7}}b-c shows the robustness of our method against the addition of series resistances to the resistance of the device. Adding a fictive resistance, $R_\mathrm{add}$, to the real $R_\mathrm{s}$ of our 2-terminal measured device before fitting $R_\mathrm{s}$, shows a linear relationship between the fitted resistance, $R_\mathrm{fit}$, and  $R_\mathrm{add}.$  
\begin{suppfigure}[h]
\centering
\includegraphics[width=1\linewidth]{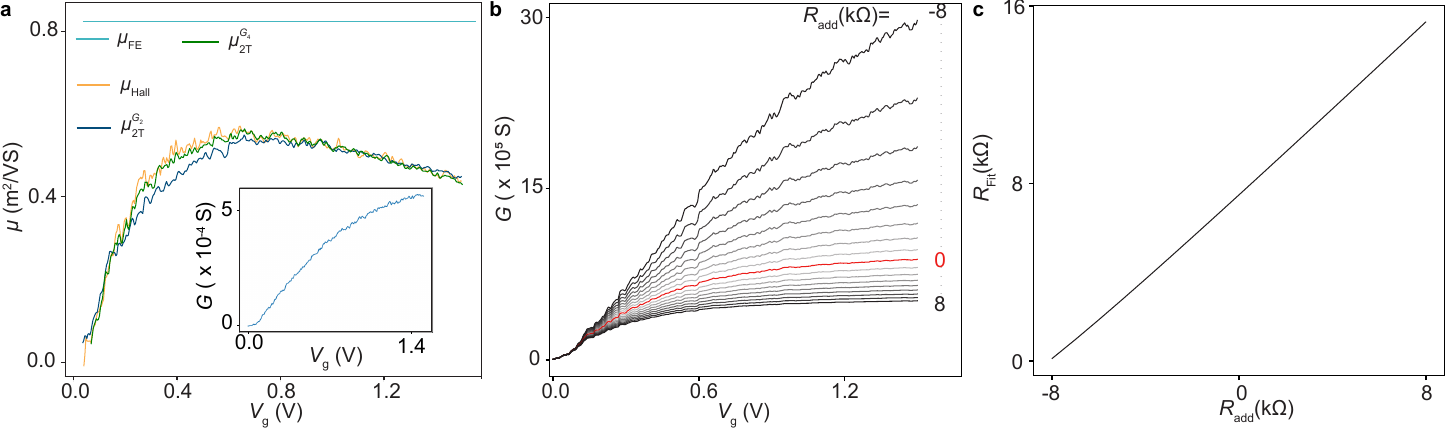}
\caption{\textbf{a} shows $\mu_\mathrm{2T}^{G_4}$, $\mu_\mathrm{2T}^{G_2}$ and $\mu_\mathrm{FE}$ compared to $\mu_\mathrm{Hall}$. The inset shows the $G$ vs. $V_\mathrm{g}$ of the 4-terminal measurement. \textbf{b} shows $G$ vs. $V_\mathrm{g}$ after adding a fictive resistances, $R_\mathrm{add}$, to the real  $R_\mathrm{s}$. \textbf{c} shows the linear relationship between $R_\mathrm{add}$ and the fitted $R_\mathrm{s}$ \label{fig:suppFig7} }
\end{suppfigure}
\subsubsection*{S8. Statistical analysis of all 256 devices at T = 100K}
Figure \textbf{S\ref{fig:suppFig8}} 
shows the same analysis as performed in Fig. 2 conducted at 100K. The trend observed is that $\mu^\mathrm{peak}_\mathrm{FE} > \mu^\mathrm{peak}_\mathrm{2T}$. While there are a few instances where the opposite occurs (e.g. highlighted data points in \textbf{f-g}), this is not an artifact of our model but instead attributed to the measurement range of $G(V_\mathrm{g})$ being too short, causing the FE-fit to place more weight on data points where the true mobility is low. An example of the inconsistency of the FE-fit is illustrated in \textbf{i-j}. In this context, \textbf{i} represents the mobilities of the red data point in \textbf{f}, while \textbf{j} corresponds to the green data point. For reference, the mean of the $\mu_\mathrm{2T}$ is plotted, which would match $\mu_\mathrm{FE}$ if $V_\mathrm{0}=V_\mathrm{th}$. However, this is not the case as shown in \textbf{h}.
\begin{suppfigure}[h]
\includegraphics[width=0.95\linewidth]{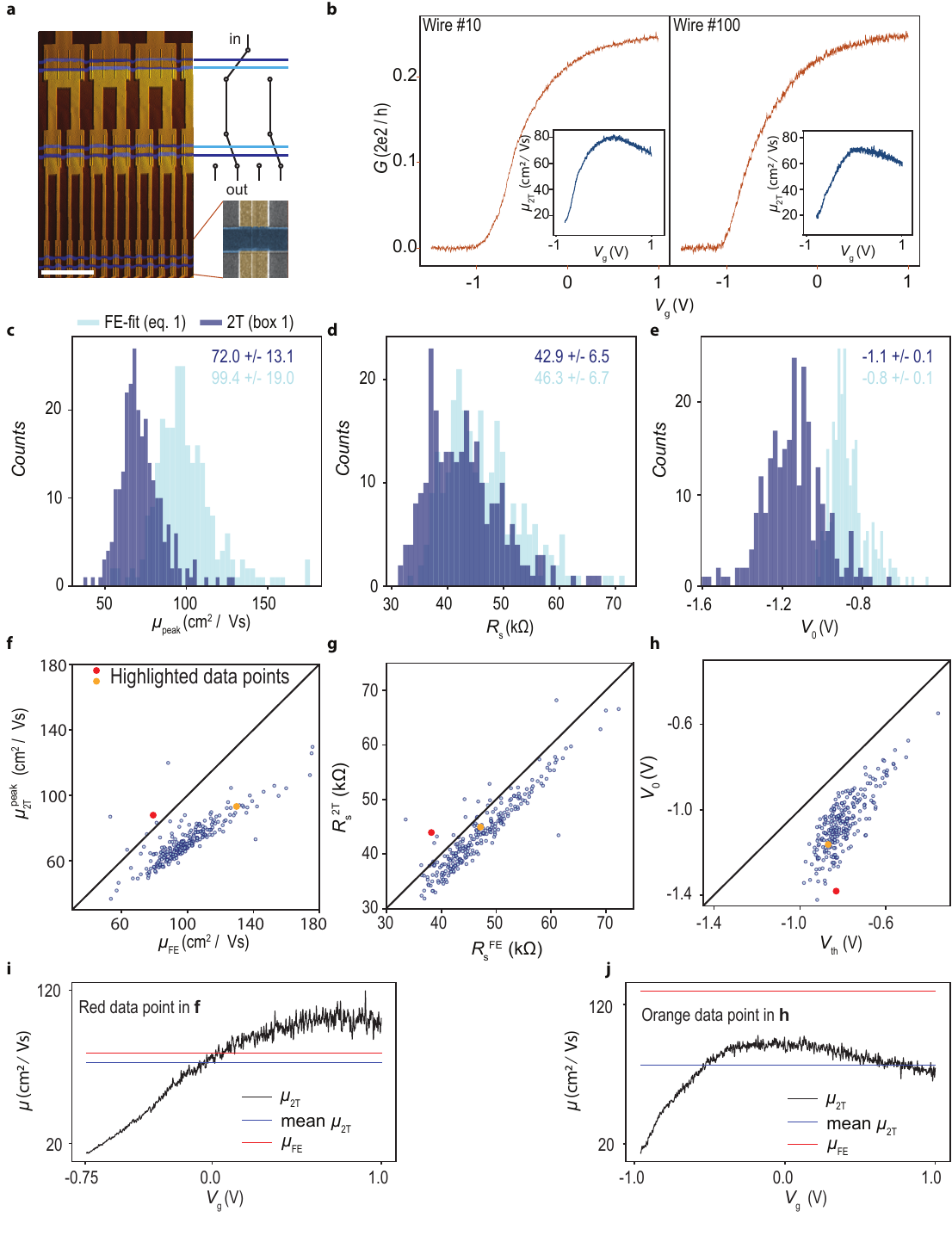}
\caption{\textbf{a-h} same analysis as in Fig. 2 of the main manuscript. Comparison of FE-fit and new fit methods at 100~K, showing that $\mu^\mathrm{peak}_\mathrm{FE} > \mu^\mathrm{peak}_\mathrm{2T}$ in most cases. Exceptions (highlighted in \textbf{f--h}) arise from limited $G(V_\mathrm{g})$ measurement range, which biases the FE-fit towards a lower value. Panels \textbf{i--j} illustrate such inconsistencies using the red and orange data points from \textbf{f}. The mean $\mu_\mathrm{2T}$ is plotted for reference and would equal $\mu_\mathrm{FE}$ only if $V_0 = V_\mathrm{th}$, which is not generally observed (see \textbf{h}).
 \label{fig:suppFig8} }
\end{suppfigure}

\subsubsection*{S9. Raw data MPX} Figure \textbf{S\ref{fig:suppFig9}} presents two-terminal $G$ vs. $V_\mathrm{g}$ for 256 SAG-based MPX FET devices, along with corresponding FE fits. These measurements form the basis for the statistical analysis also shown in \textbf{Fig.~S\ref{fig:suppFig9}}. The devices are identical to those displayed in the main text \textbf{Fig.~2}.  At $T \sim 20$~mK, only every other device was measured, resulting in data from 128 devices being shown in this specific dataset.

\begin{suppfigure}[h]
\centering
\includegraphics[width=0.95\linewidth]{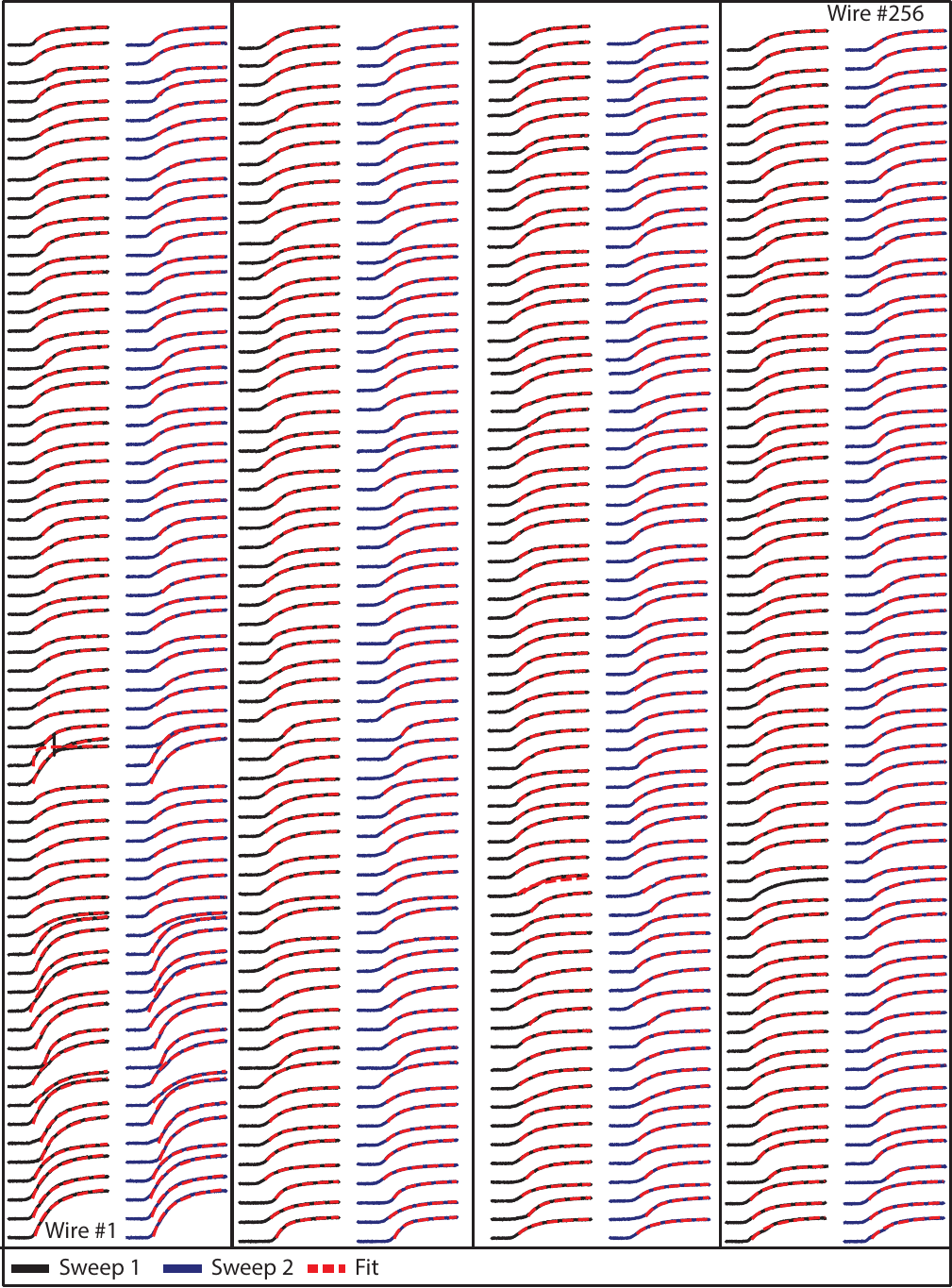}
\caption{Waterfall plot of G(V$_\mathrm{g}$)-traces of 256 wires ordered from lower left corner to upper right corner with fit to Eq. Gate was swept in the forward direction. Sweep 1 and sweep 2 are shifted by 3 volts for clarity. The same wires were also swept in the backwards direction \label{fig:suppFig9} }
\end{suppfigure}
%
%

\clearpage
\section*{Section 2}
\noindent This overview section examines models, foundational to the paper, used to extract mobilities from transport measurements, all beginning with the Drude assumption:
\begin{equation}
    \sigma = e(\mu_\mathrm{n}n + \mu_\mathrm{p}p)
\end{equation}

\noindent\textbf{The Hall Mobility}:
\newline We define the Hall resistance as:
\begin{equation}
    R_\mathrm{H} = \frac{p\mu_\mathrm{p}^2 - n\mu_\mathrm{n}^2}{e(p\mu_\mathrm{p} + n\mu_\mathrm{n})^2} = \frac{p - nb^2}{e(p + nb)^2}, \quad \text{where} \quad b = \frac{\mu_\mathrm{n}}{\mu_\mathrm{p}}
\end{equation}
and the Hall mobility as\cite{doi:https://doi.org/10.1002/0471749095.ch8}:
\begin{equation}
    \mu_\mathrm{H} = \lvert R_\mathrm{H} \rvert \sigma \label{Hall_mobility}
\end{equation}
Assuming only one type of carrier, e.g., $n$-type Drude, is recovered from Eq. \ref{Hall_mobility}:
\begin{equation}
    \mu_\mathrm{H} = \mu_\mathrm{n} = \frac{1}{ne} \sigma \equiv \mu_\mathrm{2T} \label{Hall_mobility}
\end{equation}

\noindent\textbf{The effective two-terminal extracted mobility}:
\newline The total current, $I$, in the wire is given as the sum of a drifting and diffusion current:
\begin{equation}
    I = GV_\mathrm{DS} - f(V_\mathrm{DS},T)
\end{equation}
\noindent Where $f$ is the diffusion term that depends on $V_\mathrm{DS}$ and temperature, $T$, and is assumed to be zero for small $V_\mathrm{DS}$, e.g., $V_\mathrm{DS} \ll V_\mathrm{g}$ since it is dependent on the distribution of different types of carriers. We use Drude equation and the relation, $G = \sigma\frac{W}{L}$, where $W,L$ are the width and length of the n-type-carrier-device and get: 
\begin{equation}
    G (V_\mathrm{g} = \mathrm{fixed}) = \left. \frac{\partial I}{\partial V_\mathrm{DS}} \right|_{V_\mathrm{g}} = \text{constant} = ne\mu_\mathrm{2T} \frac{W}{L} \label{G}
\end{equation}
\noindent Next, we assume that the oxide capacitance per unit area can be approximated by a constant \textit{effective} oxide capacitance, the charge density is given by:
\begin{equation}
    ne = C_\mathrm{eff}(V_\mathrm{g} -V_\mathrm{0})
\end{equation}
Substituting this into Eq. \ref{G}:
\begin{equation}
    G(V_\mathrm{g}) = C_\mathrm{eff}(V_\mathrm{g} -V_\mathrm{0})\mu_\mathrm{2T} \frac{W}{L} \label{G2}
\end{equation}
and solving for $\mu_\mathrm{2T}$:
\begin{equation}
    \mu_\mathrm{2T} = \frac{G}{C_\mathrm{eff}(V_\mathrm{g} -V_\mathrm{0})} \frac{L}{W} \label{u_eff}
\end{equation}
Note that the $V_\mathrm{0}$, used here is usually approximated by the ill-defined $V_\mathrm{th}$. Here we define $V_\mathrm{0}$ as the zero-density point for the extrapolated linear density assuming a constant capacitance model. 

\noindent\textbf{The field effect mobility}:\newline Assumption:
\begin{equation}
    g = \left. \frac{\partial I}{\partial V_\mathrm{g}} \right|_{V_\mathrm{DS}} = \frac{\partial G}{\partial V_\mathrm{g} \partial V_\mathrm{DS}} = \text{constant} \Leftrightarrow \frac{\partial G}{\partial V_\mathrm{g}} = \text{constant}\label{g}
\end{equation}
Substituting expression for $G$ in Eq. \ref{G2} into Eq. \ref{g}:
\begin{equation}
    \frac{\partial G}{\partial V_\mathrm{g}} (C_\mathrm{eff}(V_\mathrm{g} -V_\mathrm{0})\mu_\mathrm{2T} \frac{W}{L}) =\text{constant}\label{dgdvg}
\end{equation}
\begin{equation}
    \frac{\partial G}{\partial V_\mathrm{g}} = C_\mathrm{eff} \frac{W}{L} \left(\mu_\mathrm{2T} + (V_\mathrm{g} -V_\mathrm{0})\frac{d\mu_\mathrm{2T}}{dV_\mathrm{g}} \right)\label{dgdvg2}
\end{equation}
For Eq. \ref{dgdvg} to be true, it implies that Eq. \ref{dgdvg2} that $\mu_\mathrm{2T}$ is independent of $V_\mathrm{g}$.

\begin{figure*}[h]
\centering
\textbf{Comparison of Hall mobility and field effect mobility}\par\medskip
\includegraphics[width=0.990\linewidth]{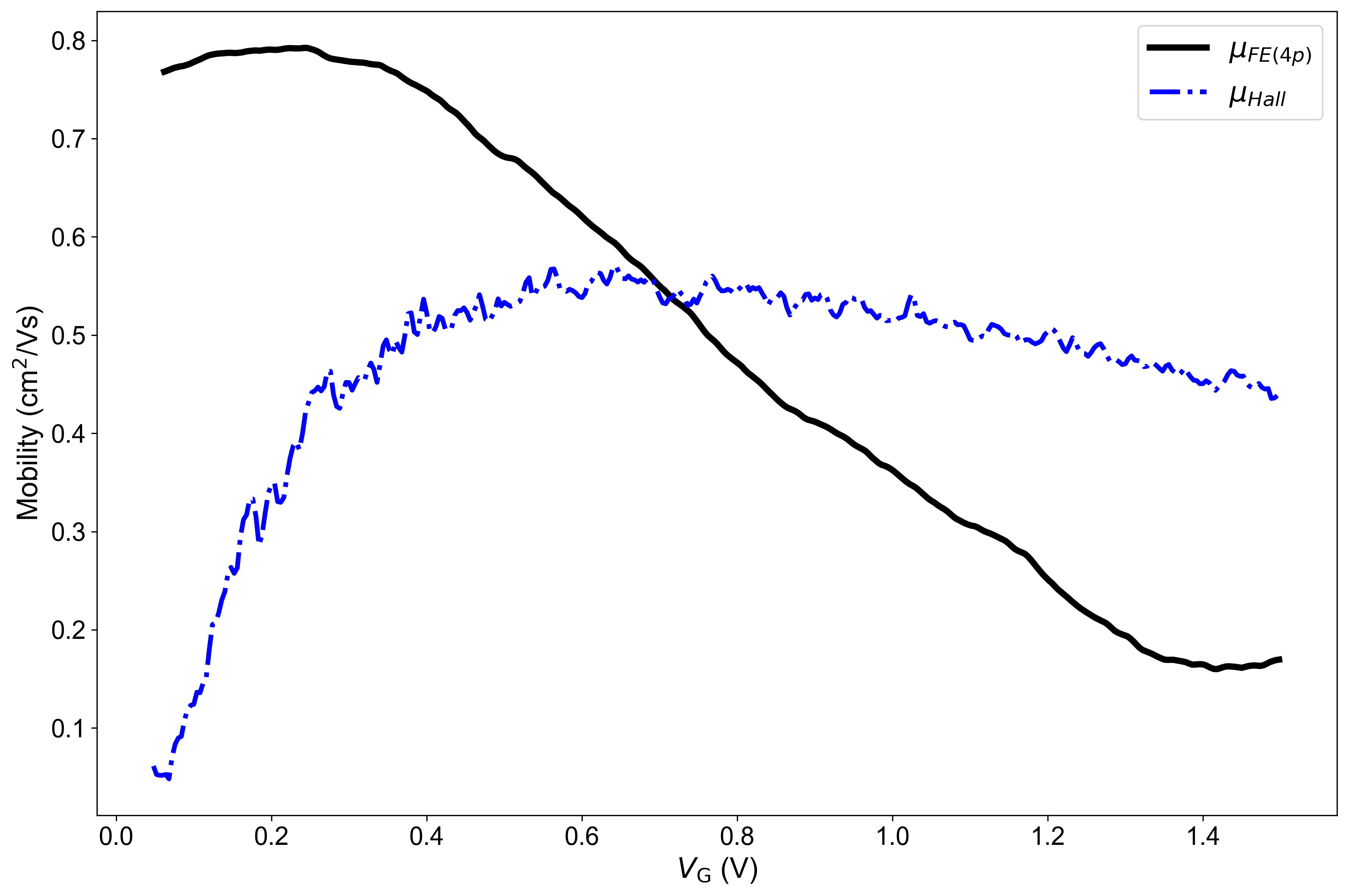}
\caption*{ Since $\mu_\mathrm{2T}$ is not independent of $V_\mathrm{g}$ it becomes clear from Eq. \ref{dgdvg2} why $\mu_\mathrm{FE} = g \frac{L}{C_\mathrm{eff}\cdot W}$ overestimates the mobility at low voltages where the $\mu_\mathrm{Hall}$ has a positive slope and vice versa. The two mobilities coincide at the turning point of the slope of $\mu_\mathrm{Hall}$. The $\mu_\mathrm{FE}$ was extracted from the 4-terminal measured conductance to avoid complications from a series resistance. \label{fig:compare_µ} }
\end{figure*}

\begin{figure*}[h]
\centering
\includegraphics[width=0.990\linewidth]{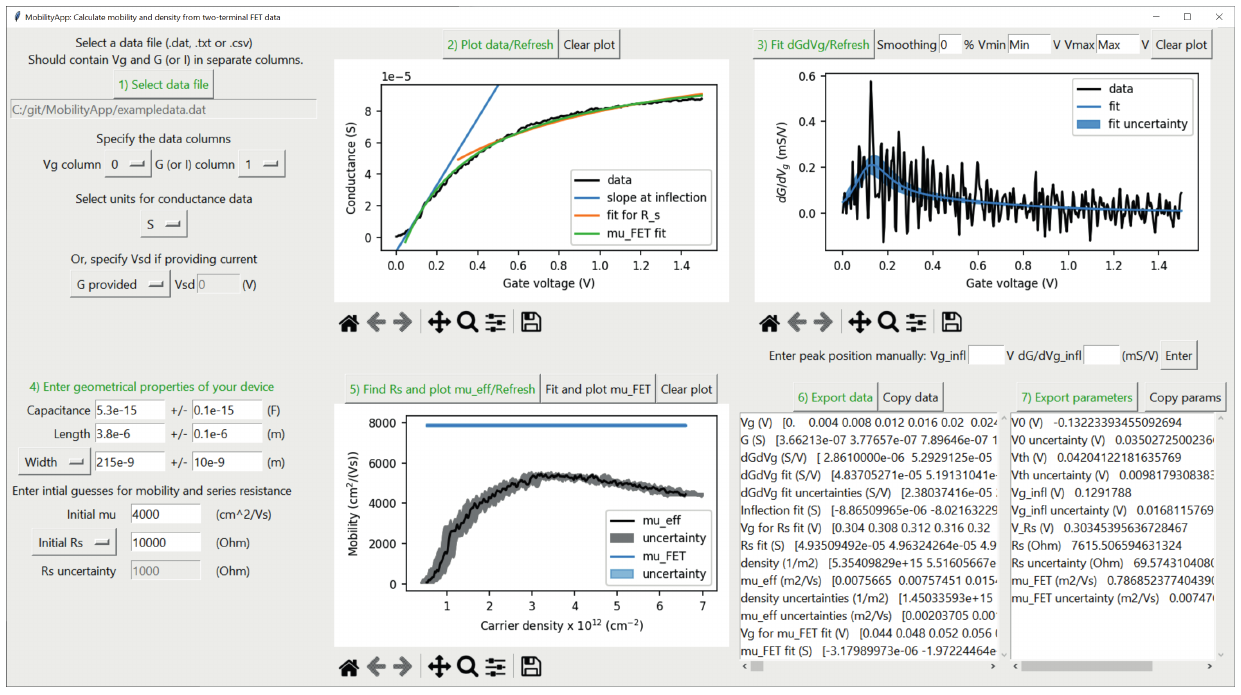}
\caption*{Screenshot of the graphical user interface of the MobilityApp. \label{fig:app}}
\end{figure*}

\section{MobilityApp}
The procedure used to transform $G(V_\mathrm{g})$ traces into $\mu(n)$ is freely available and accessible via the source code and GUI (See screenshot) available at github.com/djcarrad/mobilityapp. Windows users interested in fitting a few traces can download the GUI packaged into an .exe at github.com/djcarrad/MobilityApp/releases. General access to the fitting functions and batch processing is available for python users on all operating systems by downloading the mobilityapp package through pip (i.e. pip install mobilityapp). The example jupyter notebook provides a documentation of how to use the code and an explanation of each function. The GUI is also available directly from the command line after installation by calling mobilityapp.

\section{Uncertainty Propagation}
Uncertainties in $n(V_\mathrm{g})$ and $\mu_\mathrm{2T}(V_\mathrm{g})$ were calculated by propagating the uncertainties of all constituent variables as follows.
\begin{enumerate}
    \item The uncertainty in $V_\mathrm{infl}$, denoted $\delta V_\mathrm{infl}$, was obtained from the output of the asymmetric lorentzian fit.
    \item The uncertainty in the peak value of $\partial G/\partial V_\mathrm{g}$ was estimated by combining the uncertainties from the asymmetric lorentzian parameters $A$, $c$, $a$ and $V_\mathrm{infl}$ to give minimum and maximum $\partial G/\partial V_\mathrm{g}(V_\mathrm{g})$ functions. The final value $\delta (\partial G/\partial V_\mathrm{g} |_\mathrm{max})$ was calculated as the difference in peak values of the maximum and minimum functions.
    \item The uncertainty in $G$ around the inflection point, $\delta G_\mathrm{infl}$, was extracted from the measured data, $G(V_\mathrm{infl}+\delta V_{infl}) - G(V_\mathrm{infl}-\delta V_{infl})$.
    \item The uncertainty in $V_\mathrm{th}=V_\mathrm{infl}-(G_\mathrm{infl}/(\partial G/\partial V_\mathrm{g} |_\mathrm{max}))$, i.e., where the fit line in main text box panel a intercepts $V_\mathrm{g}$, was calculated as the difference in minimum and maximum intercept values, based on $\delta V_\mathrm{infl}$, $\delta (\partial G/\partial V_\mathrm{g} |_\mathrm{max})$ and $\delta G_\mathrm{infl}$.
    \item The uncertainty in $V_\mathrm{0} = V_\mathrm{th}-2(V_\mathrm{0}-V_\mathrm{th})$ was given by $\delta V_\mathrm{0} = \sqrt{(\delta V_\mathrm{th})^2-(2\delta V_\mathrm{infl})^2}$.
    \item The uncertainty in series resistance, $\delta R_\mathrm{S}$, was given by the fit to the Drude model (see main text box panel a).
    \item The uncertainties in $n(V_\mathrm{g})=C^i(V_\mathrm{g}-V_\mathrm{0})/e$ were calculated from $\delta n (V_\mathrm{g}) = \frac{1}{e}\sqrt{(\delta C^i(V_\mathrm{g}-V_\mathrm{0}))^2 + (C^i\delta V_\mathrm{0})^2}$, where $\delta C^i$ is the uncertainty in capacitance per area obtained from COMSOL simulations (see section XX).
    \item Finally, the uncertainties in $\mu_\mathrm{2T} (V_\mathrm{g})$ were calculated by computing the partial derivatives $\partial \mu_\mathrm{2T}/\partial x$ for $x = L, C, V_\mathrm{0},R_\mathrm{S}$ and using $\delta \mu_\mathrm{2T}(V_\mathrm{g}) = \sqrt{(\frac{\partial \mu_\mathrm{2T}}{\partial L}\delta L )^2+(\frac{\partial \mu_\mathrm{2T}}{\partial C}\delta C)^2+(\frac{\partial \mu_\mathrm{2T}}{\partial V_\mathrm{0}}\delta V_\mathrm{0})^2+(\frac{\partial \mu_\mathrm{2T}}{\partial R_\mathrm{S}}\delta R_\mathrm{S})^2}$ where $\delta L$ is the uncertainty in the device length as estimated from SEM imaging.
\end{enumerate}

The python code used to calculate uncertainties is freely available as part of the MobilityApp package.

\bibliography{bib}